\documentclass[reprint,english,aps,prl, twocolumn, amsmath, amssymb, showpacs, showkeys, superscriptaddress]{revtex4}
\usepackage[T1]{fontenc}
\usepackage{subfigure}
\usepackage[latin1]{inputenc}
\usepackage{graphicx}
\usepackage{epsfig}
\usepackage{prettyref}
\usepackage{babel}
\usepackage{caption}
\makeatother

\begin{document}

\title{Large linear magnetoresistance from neutral defects in  Bi$_2$Se$_3$}

\author{Devendra Kumar}
\email{deveniit@gmail.com}
\affiliation{UGC-DAE Consortium for Scientific Research, University Campus, Khandwa Road, Indore-452001, India}

\author{Archana Lakhani}
\email{archnalakhani@gmail.com}
\affiliation{UGC-DAE Consortium for Scientific Research, University Campus, Khandwa Road, Indore-452001, India}

\begin{abstract}
The chalcogenide Bi$_2$Se$_3$ can attain the three dimensional (3D) Dirac semimetal state under the influence of strain and microstrain. Here we report the presnece of large linear magnetoresistance in such a  Bi$_2$Se$_3$  crystal. The magnetoresistance has quadratic form at low fields which crossovers to linear above 4~T. The temperature dependence of magnetoresistance scales with carrier mobility and the crossover field scales with inverse of mobility. Our analysis suggest that the linear magnetoresistance in our system has a classical origin and arises from the scattering of high mobility 3D Dirac electrons from crystalline inhomogeneities. We observe that the charged selenium vacancies are strongly screened by high mobility Dirac electrons and the neutral crystalline defects are the main scattering center for transport mechanism. Our analysis suggests that both the resistivity and the magnetoresistance have their origin in scattering of charge carriers from neutral defects.
\end{abstract}


\maketitle

\section{Introduction}The topological insulators are novel materials having spin polarized Dirac electrons at the conducting surface  and an insulating bulk~\cite{Hasan, Qi}. Of these, the chalcogenide Bi$_2$Se$_3$ is the most appealing because of simple gapless Dirac cone at the surface and the large topologically non-trival gap of 0.3eV  between the bulk bands~\cite{Zhang}.  The bulk band gap of topological insulators can be reduced on varying the pressure, strain and strength of spin orbit coupling~\cite{ Young, Liu1, Xu, Sato, Brahlek, Vanderbilt}. The change in the topological band gap can manifest a zero gap state giving three dimensional (3D) Dirac point with linear dispersion in all directions, known as 3D Dirac semimetal or 3D Dirac fermion metal state, where a topological transition from topological to trivial band insulator occurs~\cite{Young1, Yang, He1, Murakami}. Such 3D Dirac fermion metal state at the topological transition have been observed to occur on varying the strength of spin orbit coupling through doping such as in TlBi(S$_{1-x}$Se$_x$)$_2$~\cite{Xu, Novak}, Pb$_{1-x}$Sn$_x$Se~\cite{Zeljkovic}, Bi$_{1-x}$Sb$_x$~\cite{Teo, Kim} or on varying the lattice parameter through strain and microstrain in Bi$_2$Se$_3$~\cite{Liu, Devendra}. The 3D Dirac semimetal state has also been realized in materials with certain crystallgraphic symmetries that allows 3D Dirac point at high symmetry points of Brillouin zone~\cite{Young2}, for example, in Na$_3$Bi~\cite{Liu2} and Cd$_3$As$_2$~\cite{Liu3}.

The massless Dirac fermions in two dimensional (2D) surface states of topological insulators and in 3D Dirac semimetals exhibit  interesting magnetotransport properties such as large linear magnetoresistance (MR)~\cite{Wang2, Novak, Narayanan}, non-trival Berry phase in Shubnikov-de Haas (SdH) oscillations~\cite{Analytis1, Yan1, Novak, Narayanan, He1}, and weak antlilocalization (WAL)~\cite{Kim1, Zhao}. The linear magnetoresistance in topological insulators is observed in thin films, nanoplates/nanoribbons of Bi$_2$Se$_3$~\cite{Tang, He, Yan, Hongtao}, nanosheets and crystals of Bi$_2$Te$_3$~\cite{Wang2, Barua, Qu1},  where the surface state contribution dominates the overall transport of the system. The linear MR in nanosheets of Bi$_2$Te$_3$~\cite{Wang2} and nanoribbons of Bi$_2$Se$_3$~\cite{Tang} has been attributed to Abrikosov theory of quantum linear magnetoresistance proposed for zero gap materials with linear dispersion. The linear MR in nanoplates and thin films of Bi$_2$Se$_3$ have also been  attributed to mobility fluctuations due to inhomogeneities and therefore suggesting a classical origin~\cite{He, Yan}. For 3D Dirac semimetals, the large linear MR is observed in bulk crystals of materials existing at the topological transition e.g. TlBiSSe~\cite{Novak} as well as in materials having the crystallographic symmetry e.g. Cd$_3$As$_2$~\cite{Narayanan} and Na$_3$Bi~\cite{Xiong}.

In this work we present the magnetotransport study on a Bi$_2$Se$_3$ crystal exhibiting signatures of 3D Dirac fermion metal state due to topological phase transition caused by microstrain~\cite{Devendra}. The magnetoresistance grows with the applied field and becomes linear above 4~T.  The linear MR is non-saturating, temperature dependent, has bulk origin, and scales well with the carrier mobility. The magnetoresistance values are large in comparison to the previously reported values for bulk Bi$_2$Se$_3$~\cite{Cao, Eto, Devidas}  The analysis of magnetotransport data suggest that the linear MR in our system arises from the  fluctuations in carrier mobility due to scattering from low mobility inhomogeneity islands. Our results indicate that charged selenium vacancies are strongly screened by the high mobility 3D Dirac electrons and the neutral defects are the main source of electron scattering in the system.

\section{Experimental Details}
Bi$_2$Se$_3$ single crystal sample used in the present study is the same one used in reference~\cite{Devendra}. The sample used for resistivity and Hall measurements is in a thin rectangular bar shape. Resistivity and Hall measurements were performed on a 9T PPMS AC Transport system (Quantum Design) using standard four probe ac and five probe ac technique respectively. The angle dependent  magnetoresistance measurements were carried on a high resolution rotator probe of PPMS.

\begin{figure}[t]
\begin{centering}
\includegraphics[width=0.8\columnwidth]{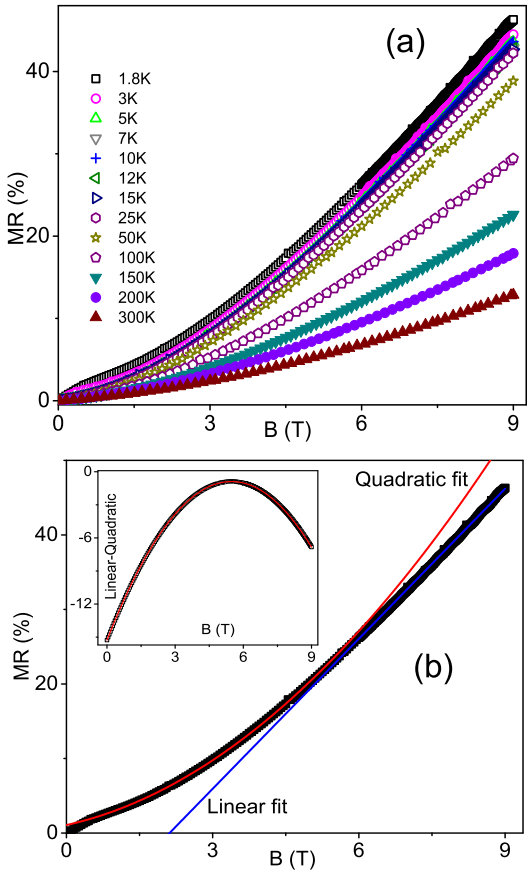}
\par\end{centering}
\caption{(a) The field dependence ($B||C_3$ axis) of magnetoresistance (MR) at different temperatures. (b) The MR data at 1.8~K. The low field data is fitted with quadratic function while the high field data is fitted with the linear function to demonstrate their respective field dependence. The field variation of the difference of linear and quadratic function is plotted in the inset.}
\label{fig: TSdH}
\end{figure}

\begin{figure}[t]
\begin{centering}
\includegraphics[width=0.8\columnwidth]{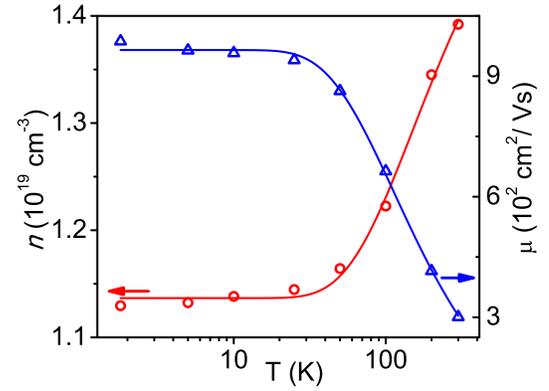}
\par\end{centering}
\caption{The temperature variation of carrier concentration $n$ (open circles) and mobility $\mu$ (open triangles). The solid lines represent the least square fitting of equation~\ref{eq:Ar} to $n$ and equation~\ref{eq:mu} to $\mu$. }
\label{fig: carriermob}
\end{figure}

\section{Results and Discussion}

\subsection{Linear magnetoresistance}
Figure~\ref{fig: TSdH}~(a) displays the magnetoresistance MR=($\rho(B)$-$\rho(0)$)/$\rho(0)$ as a function of applied field ($B||C_3$ axis) upto 9~T at various temperatures. The magnetoresistance is large, $\sim$50\% at 1.8~K, exhibits a weak temperature dependence upto $\sim$25~K, and thereafter decreases strongly on increasing the temperature. The magnetoresistance shows a quadratic field dependence at low fields which crossover to linear dependence at high magnetic fields.  The crossover field ($B_C$) is determined by first fitting the low field magnetoresistance to quadratic and high field magnetoresistance to linear function, and then $B_C$ is obtained as the field at which the absolute value of the difference between these linear and quadratic functions is minimum. See figure~\ref{fig: TSdH}~(b) and its inset. The existence of linear magneto-resistance above $B_C$ can not be understood on the basis of Lorentz deflection of carriers as the Lorentz deflection predicts a quadratic magneto-resistance which may saturate in case of closed orbits.

\subsection{Quantum linear magnetoresistance: Feasibility}
The Bi$_2$Se$_3$ single crystal under study exhibits signature of 3D Dirac fermion metal state~\cite{Devendra}. In this state, the Fermi energy of the system $E_F$$\sim$336~meV above the Dirac point. The 3D Dirac fermions have an effective mass $m^*$=0.15$m_e$ which gives energy spacing between two consecutive Landau levels $\Delta E$=$\hbar\omega_c\sim$3.9 meV for magnetic field $B$=5~T, where $\omega_c$=$eB/m^*$ is the cyclotron frequency.  The non-saturating linear magnetoresistance at high fields can arise from quantum as well as classical effects.  Abrikosov~\cite{Abrikosov, Abrikosov1} proposed a quantum interpretation of high field linear magnetoresistance for materials processing a gapless linear energy dispersion in ultra quantum limit. In ultra quantum limit, only the first landau level is filled and this happens when $\hbar\omega_c$>$E_F$, and $\hbar\omega_c$, $E_F \gg k_BT$. In our system, the linear magnetoresistance at low temperature sets in $\sim$5~T and therefore the condition $\hbar\omega_c$>$E_F$ is not satisfied. At this field, the $E_F$ lies around the 86th Landau level which is far from the quantum limit. This suggest that the high field linear magnetoresistance in our system is not due to the quantum effects. The quantum model predicts a temperature independent magnetoresistance till the thermal energy is smaller than the $\hbar\omega_c$ and $E_F$. In our case, $\hbar\omega_c$ for 9~T is $\sim$ 81~K whereas the magnetoresistance starts dropping strongly above 25~K (see Figure~\ref{fig: TSdH}~(a)) which further contradicts the validity of quantum model.

\subsection{Neutral defects: the dominant scatterers}
Figure~\ref{fig: carriermob} shows the temperature variation of carrier concentration ($n$) and electron mobility ($\mu$) estimated from the Hall resistivity and the ratio of Hall coefficient and longitudinal resistivity respectively. The carrier concentration ($n$) increases on increasing the temperature and fits well with the Arrhenius's law:
\begin{equation}
n(T)=n_0(1+\textrm{exp}(-\Delta/k_BT)),\label{eq:Ar}
\end{equation}
where $n_0$ is the carrier concentration at T=0~K, $\Delta$ is the activation energy, and $k_B$ is the Boltzmann constant. The fitting gives $n_0$=1.13$\times$10$^{19}$cm$^{-3}$ and $\Delta$=13~meV. Since  bulk Dirac electrons, the intrinsic charge carriers of a 3D Dirac fermion metal are not expected to follow a Arrhenius's type behavior, the thermal activation of charge carriers is possibly coming from some charge traps of our system~\cite{Farmer, Li, Fang}. In contrast to carrier concentration, the carrier mobility ($\mu$) exhibits a strong decay on enhancing the temperature. See figure~\ref{fig: carriermob}. The temperature variation of mobility does not follow the power law dependence (T$^{3/2}$ for impurity scattering and T$^{-3/2}$ for phonon scattering) of normal metals/semiconductors. For Dirac semimetals, at temperatures lower than Fermi temperature $T_F$, $\mu(T)\propto T^2$ for scattering from Coulomb defects while
\begin{equation}
\mu(T)=\mu(0)(1-\textrm{exp}(-T_F/T)),\label{eq:mu}
\end{equation}
for scattering from neutral defects~\cite{Sarma}. The mobility due to phonon scattering is $\propto T^{-1}$ in Bloch-G\"{r}uneisen (BG) regime i.e. for $T<T_{BG}$ where $T_{BG}=2v_{ph}k_F/k_B$, $v_{ph}$ is phonon velocity, $k_F$ is the Fermi wave vector and $k_B$ is Boltzman constant. For Bi$_2$Se$_3$ $v_{ph}$ =2900 m/s which gives $T_{BG}\approx$~1230~K. The phonon contribution of mobility is generally weak at low temperatures and is difficult to detect experimentally. Therefore at low temperatures, the overall mobility of a Dirac semimetal is determined by the scattering from charged and neutral defects.  The temperature dependence of mobility shown in figure~\ref{fig: carriermob} fits well with equation~\ref{eq:mu} suggesting that the scattering from neutral defects dominates the carrier transport mechanism. This also signals the strong screening of Coulomb defects i.e. the selenium vacancies by the 3D Dirac electrons in our system.

\begin{figure}[]
\begin{centering}
\includegraphics[width=1.0\columnwidth]{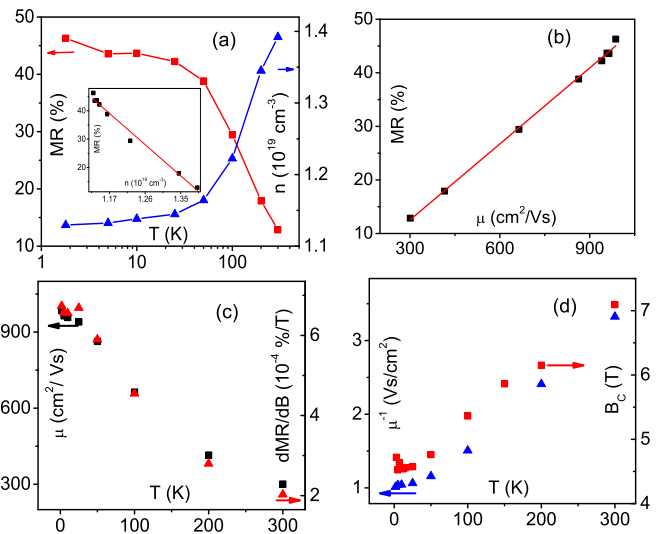}
\par\end{centering}
\caption{(a) Temperature dependence of MR at 9T (squares) and carrier concentration $n$ (triangles). The inset shows the scaling of MR with the temperature induced variation in $n$. The straight line is the linear fit to MR versus $n$. (b) The MR (at 9T) versus carrier mobility $\mu$ on temperature variation. The straight line represents the linear fit to data. (c) Temperature dependence of carrier mobility $\mu$ (squares) and the slope of linear MR, dMR/dB (triangles). (d) Evolution of inverse of carrier mobility $\mu^{-1}$ (triangles) and crossover field $B_C$ (squares) with temperature.}
\label{fig: Scaling1.eps}
\end{figure}

\subsection{Classical linear magnetoresistance: Feasibility}
The other mechanisms for the existence of linear magetoresistance are classical in nature and applicable to inhomogeneous systems with high carrier mobility. One such model from Herring~\cite{Herring} predicts that the small fluctuations in local conductivity can lead to linear magnetoresistance at extremely high fields in the limit of weak disorder. For strong disorder limit, Parish and Littlewood (PL)~\cite{Parish, Parish1} proposed that the distortion in current path due to inhomogeneous distribution of carrier concentration and mobility can give rise to linear magnetoresistance. The PL model uses a random resistor network to simulate an inhomogeneous system where regions of different local carrier concentration and mobility are represented by a unique four terminal resistor. Each four terminal resistor accounts for the resistive as well as Hall voltage of a particular region. The magnetoresistance in PL model strongly dependents on the average mobility ($\langle\mu\rangle$) and the width of mobility disorder ($\Delta\mu$). At high fields, $MR\propto$$\langle\mu\rangle$ for $\Delta\mu$/$\langle\mu\rangle$<1 and $MR\propto\Delta\mu$ for $\Delta\mu$/$\langle\mu\rangle$>1. The crossover field ($B_c$) for quadratic to linear MR is $\propto\mu^{-1}$ for $\Delta\mu$/$\langle\mu\rangle$<1 and $\propto$($\Delta\mu)$$^{-1}$ for $\Delta\mu$/$\langle\mu\rangle$>1. Therefore, the behavior of MR with carrier mobility and other parameters will be useful in understanding the mechanism of linear magnetoresistance.

Figure~\ref{fig: Scaling1.eps}~(a) compares the temperature dependence of magnetoresistance (at 9~T) and carrier concentration.  The magnetoresistance decreases while the carrier concentration increases on increasing the temperature. The  MR drops linearly as  $n$ increases (see inset of Figure~\ref{fig: Scaling1.eps}~(a)), but the magnitude of change in MR and $n$ are of different orders suggesting that the change in MR is not determined by the thermally activated change in carrier concentration $n$.  Figure~\ref{fig: Scaling1.eps}~(b) displays the variation of MR with carrier mobility on changing the temperature. The size of MR grows linearly with carrier mobility implying that $MR\propto\mu$. The dependency of linear MR on carrier mobility is further demonstrated by the temperature dependence of carrier mobility and slope of linear part of magnetoresistance ($dMR/dB$). See Figure~\ref{fig: Scaling1.eps}~(c). $dMR/dB$ is obtained by linear fitting of MR at high fields. A remarkable similarity between the temperature evolution of $dMR/dB$ and carrier mobility  confirms unambiguously that the $MR\propto\mu$. Figure~\ref{fig: Scaling1.eps}~(d) exhibits the the temperature dependence of crossover field ($B_c$) and compares it with that of inverse mobility ($\mu^{-1}$). $B_c$ increases on enhancing the temperature and its temperature dependence scales well with that of $\mu^{-1}$.   This indicates that $B_c\propto\mu^{-1}$. The dependency of MR and $B_c$ on carrier mobility i.e. $MR\propto\mu$ and $B_c\propto\mu^{-1}$ confirms that the linear MR is classical and can be understood on the basis of PL model for narrow mobility distribution $\Delta\mu$/$\langle\mu\rangle$<1~\cite{Parish}. The linear magnetoresistance in inhomogeneous systems follow the PL model in a variety of materials such as Ag$_{2+\delta}$Se and  Ag$_{2+\delta}$Te~\cite{Hu}, epitaxial graphite~\cite{Aamir} and graphene~\cite{Wang1}, topological insulator YPdBi~\cite{Wang}, and the Dirac semimetal Cd$_3$As$_2$~\cite{Narayanan}.

\begin{figure}[!t]
\begin{centering}
\includegraphics[width=1.0\columnwidth]{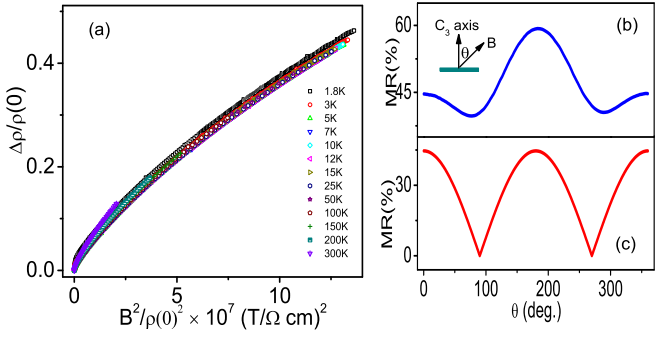}
\par\end{centering}
\caption{(a) Kohler plot exhibiting the collapse of isothermal MR on a single curve. (b) The angle dependence of MR at 8.7~T. Inset defines the tilt angle $\theta$ and (c) the $|cos\theta|$ dependence.}
\label{fig: AngleDep1}
\end{figure}

\subsection{Inhomogeneity scattering as the source of linear magnetoresistance}
In the PL model, the linear magnetoresistance in a high mobility inhomogeneous conductor arises due  to multiple scattering of electrons from low mobility islands around inhomogeneities~\cite{Kozlova}. The inhomogeneity scattering decreases the average electron velocity along the applied voltage by deflecting the electron motion perpendicular to it. For single band metal, the semiclassical theory predicts that tan$\theta_H$=$\rho_{yx}/\rho_{xx}$=$\omega_c\tau_H$ where $\theta_H$ is the Hall angle and  $\tau_H$ is the scattering time for carrier motion perpendicular to the direction of applied voltage (and magnetic field). At 1.8~K and 9~T, tan$\theta_H$=$\rho_{yx}/\rho_{xx}$= 0.44 which yields $\tau_H$$\sim$4$\times10^{-14}$s. The transport scattering time ($\tau_t$) corresponding to resistivity is $\tau_t$=$\mu_Hm^*/e$$\sim$8.4$\times10^{-14}$s.  The $\tau_H$ and $\tau_t$ are of same order suggesting that both the resistivity and magnetoresistance are governed by same scattering process. The existence of single scattering process is further analyzed by the Kohler's scaling
of isothermal MR: $\Delta \rho/\rho_0$=$F(B/\rho_0)$, where $\Delta\rho$ is the change in isothermal resistance on applying field $B$ and $\rho_0$ is the zero filed resistivity at the given temperature. The Kohler's rule is satisfied if there is a single temperature dependent scattering time at all point of Fermi surface. Figure ~\ref{fig: AngleDep1} shows the Kohler plot for isothermal MR between 1.8~K to 300~K. All MR data collapse on a single curve suggesting that a single scattering process dominates in our system.

\begin{figure}[!t]
\begin{centering}
\includegraphics[width=0.6\columnwidth]{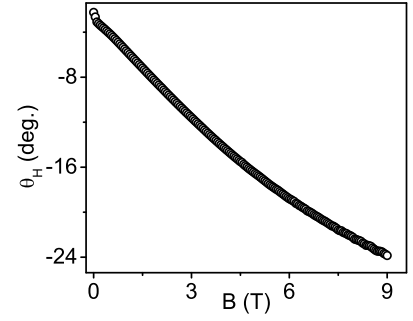}
\par\end{centering}
\caption{Field dependence of Hall angle $\theta_H$ at 1.8~K.}
\label{fig: Hallangle}
\end{figure}

To confirm whether electrons undergoing the scattering for linear MR are from surface sates or bulk, we have plotted the angle dependence of magnetoresistance at the field of $\sim$ 8.7~T in figure~\ref{fig: AngleDep1}~(b). The angle dependence of magnetoresistance differs from $|cos\theta|$ dependence shown in figure~\ref{fig: AngleDep1}~(c) suggesting that linear MR is not from surface state electrons. The linear MR in our system is different from that of nanoflakes~\cite{Yan, Tang} and thin films of Bi$_2$Se$_3$~\cite{He, Hongtao} where it originates from surface Dirac electrons. Our sample has bulk transport characteristic with 3D Dirac electrons indicating that the linear MR in our system is arising from the inhomogeneity scattering of the 3D Dirac electrons. Furthermore the deviation from $|cos\theta|$ behaviour suggest that the MR in our system is not solely governed by Hall field, as in the case of TlBiSSe with 3D Dirac electrons~\cite{Novak}, and the scattering from inhomogeneities plays a crucial role.

The possible sources of inhomogeneities in our system are the ionized  selenium vacancies and neutral crystalline defects, for example, the dislocation, mosaic spread, and misorientation. The local strain field due to inhomogeneities is compressive suggesting it arises from neutral crystalline defects~\cite{Devendra1}. For $n$=1.13$\times$10$^{19}$ cm$^{-3}$ and considering that each selenium vacancy creates two electrons, the average distance between the ionized selenium vacancies $l_{imp}$ $\simeq$ ($n/2$)$^{-1/3}$$\approx$4~nm. The Deby screening length of ionized selenium vacancies $l_\textrm{Debye}$=($\epsilon_0\epsilon_r k_BT/ne^2$)$^{1/2}$$\approx$0.3~nm, where $\epsilon_0$ is the permittivity of free space, $\epsilon_r$=100 is the relative permittivity of Bi$_2$Se$_3$~\cite{Kohler}, $k_B$ is the Boltzman constant, and $e$ is the electronic charge. The $l_{imp}$ is two orders of magnitude smaller than the electronic mean free path $l$=$v_F\tau_H$$\approx$430~nm ($v_F=6.3\times 10^5$ m/s~\cite{Devendra}) suggesting that the ionized selenium vacancies are strongly screened by the free Dirac carriers. The strong screening of ionized vacancies is also supported by the ratio of transport scattering time ($\tau_t$) and single particle scattering time ($\tau_s$); $\tau_t/\tau_s$=2.8 which is close to the values ($\approx$1.5) expected for strong coulomb screening in 3D Dirac materials~\cite{Sarma}. The screening of ionized selenium vacancies leaves the neutral defects as the other possible source of inhomogeneities. The temperature variation of mobility in our system has shown that neutral defects are the dominant source of scattering; therefore are responsible for the macroscopic spatial fluctuation in carrier mobility. This also agrees with the fact that the neutral defects are the main source of local strain field in our crystal.

Recently an alternate theory for linear magnetoresistance in 3D metals has been put forward to explain the experimental results of Dirac semimetals without evoking the Dirac nature of charge carriers~\cite{Song}. The theory is semiclassical and predicts that the gliding center (GC) diffusion causes the linear magnetoresistance. This GC diffusion can yield linear magnetoresistance even when (i) multiple Landau levels are filled and (ii) disorder is weak. To investigate whether linear magnetoresistance in our system is because of GC diffusion, we have plotted the field dependence of Hall angle at 1.8~K in figure~\ref{fig: Hallangle}. The GC diffusion magnetoresistance requires the magnetic field magnitude independent Hall angles while the Hall angle shown in figure~\ref{fig: Hallangle} varies with field. This clarifies that linear MR in our system is not because of GC diffusion.

\section{Conclusions}
In conclusion, large linear magnetoresistance having bulk origin is observed in a Bi$_2$Se$_3$ crystal. The magnetoresistance is temperature dependent and scales well with carrier mobility. Our analysis suggest that the linear magnetoresistance in our system arises from fluctuations in carrier mobility due to multiple scattering of high mobility Dirac electrons from crystalline inhomogeneties. The Kohler scaling and the agrement between $\tau_H$ and $\tau_t$ shows the presence of a single scattering process for resistivity and magnetoresistance. The temperature variation of carrier mobility and the large value of electron mean free path ($\approx$430~nm) in comparison to distance between charged selenium vacancies($\approx$4~nm) indicate the strong screening of selenium vaccanices from 3D Dirac electrons.  Our results suggest that the scattering from neutral defects is responsible for linear magnetoresistance in our system.

\section{Acknowledgements}
We acknowledge V. Ganesan and A. K. Sinha for support and encouragement.

\end{document}